
\magnification=1200
\def\ltsima{$\; \buildrel < \over \sim \;$}
\def\simlt{\lower.5ex\hbox{\ltsima}}
\def\gtsima{$\; \buildrel > \over \sim \;$}
\def\simgt{\lower.5ex\hbox{\gtsima}}

\def\spose#1{\hbox to 0pt{#1\hss}}
\def\approxlt{\mathrel{\spose{\lower 3pt\hbox{$\sim$}}
	\raise 2.0pt\hbox{$<$}}}
\def\approxgt{\mathrel{\spose{\lower 3pt\hbox{$\sim$}}
	\raise 2.0pt\hbox{$>$}}}

\centerline{\bf THE CONTRIBUTION OF THE OBSCURING TORUS}
\centerline{\bf TO THE X--RAY SPECTRUM OF SEYFERT GALAXIES:}
\centerline{\bf A TEST FOR THE UNIFICATION MODEL}
\vskip 0.5 true cm
\centerline{Gabriele Ghisellini$^1$,
Francesco Haardt$^2$ and Giorgio Matt$^{3,4}$}
\vskip 0.5 truecm

\item{1:} Osservatorio di Torino, Strada Osservatorio 20,
10025 Pino Torinese, Italy. \par
E--mail: 32065::ghisellini, ghisellini@astto2.astro.it.

\item{2:} ISAS/SISSA, via Beirut 2--4,
34013 Trieste, Italy.\par
E--mail: 38028::haardt, haardt@astmiu.mi.astro.it.

\item{3:} Institute of Astronomy, Madingley Road, Cambridge CB3 OHA.
\par
E--mail: 20003::matt, matt@mail.ast.cam.ac.uk.

\item{4:} present address: Istituto di Astrofisica Spaziale, CNR,
Via E. Fermi, 21, I--00044, Frascati, Italy
\par
E--mail: 40607::matt
\vskip 3 truecm
{\centerline{\it MNRAS, in press}}
\vskip 8 truecm
{\centerline{SISSA ref. 175/93/A}}
\bye
\vfill \eject
\vskip 0.1 truecm \noindent
{\bf ABSTRACT}\par
\noindent
The presence of an obscuring torus around
the nucleus of Seyfert galaxies, as supposed in the popular
unification scheme, can strongly modify the X--ray spectrum
for both type 1 and 2 Seyfert
galaxies. We study this problem by means of Montecarlo simulations,
finding that if the torus is Compton thick, it can scatter at small angles
a significant fraction of the nuclear radiation,
and contribute to the continuum of Seyfert 1 galaxies above $\sim$10 keV,
and to the fluorescence iron line at 6.4 keV.
At large inclination angles and for large torus column densities,
the spectrum is attenuated by photoabsorption and Compton
scattering, while the iron fluorescence line produced by the torus
can have large equivalent widths. Even
after dilution by the continuum scattered by the warm material
outside the torus, this iron line
could be strong enough to explain the ``cold" component
in the spectrum of NGC 1068. Identifying the scattering medium with the
warm absorber seen in Seyfert 1 galaxies,
we derive a lower limit of the inclination angle in this source
of about 50$^{\circ}$. Furthermore, an energy budget argument suggests
a value of about 55$^{\circ}$--70$^{\circ}$.
We stress that the complex pattern of the predicted variability
can be a powerful tool for constraining the parameters of the model,
such as the column density of the torus, its inclination and the
amount of warm scattering material. Good energy resolution
observations of the 6.4 fluorescent iron line
can indicate the relative importance of the torus line emission
with respect to cold material in the vicinity of the nuclear source,
such as an accretion disc.

\vskip 0.3 true cm
\noindent {\bf Keywords:} Radiative transfer -- Line: formation --
Polarization -- Galaxies: active -- Galaxies: Seyfert -- X--rays: galaxies
\vskip 0.5 truecm

\vfill\eject

\vskip 0.1 truecm
\noindent
{\bf 1. INTRODUCTION}
\vskip 0.7 true cm

\noindent
The popular model unifying Seyfert 1 and Seyfert 2 galaxies
(hereinafter Sey 1 and Sey 2)
assumes that the different properties of the two classes
are simply due to an inclination effect (e.g.
Lawrence \& Elvis 1982; Antonucci \& Miller 1985;
Krolik \& Begelman 1988; see also Antonucci 1993  for a review).
In this model, a geometrically and optically thick torus is supposed
to surround the active nucleus, absorbing its optical/UV
and soft X--ray radiation, including the broad emission lines.
If the inclination
of the system is such that the nucleus is hidden to us by the torus, the
source is classified as Sey 2, otherwise as Sey 1.

Several observations support the idea of inclination
being the main discriminating parameter: the presence of
broad lines in the polarized light of Sey 2 (Antonucci \& Miller
1985; Miller \& Goodrich 1990; Tran et al. 1992), which is explained as
due to the scattering into the line of sight of the hidden Sey 1 nucleus
by warm material outside the absorbing torus;
the anisotropic continuum emission implied by the ionization cones
(e.g. Pogge 1989; Tadhunter \& Tsvetanov 1989; Haniff, Ward \& Wilson 1991);
the energy and photon budget in the ionization cones
(e.g. Wilson 1992);
the large Sey 2 X--ray column densities (e.g. Awaki et al. 1991;
Mulchaey, Mushotzky \& Weaver 1992; Nandra \& Pounds 1993).

The prototypical Sey 2, NGC 1068, does not show intrinsic X--ray absorption
in the {\it Ginga} band (1-20 keV),
but this is readily explained by assuming that we are observing only the
X--rays scattered in our directions by
the same warm material scattering (and polarizing) the broad lines.
We would therefore observe in this source only a tiny fraction
of the entire luminosity.
This hypothesis is confirmed by recent, moderate energy resolution BBXRT data
(Marshall et al. 1993).

Present X--ray data of Sey 2 indicate absorbing column densities
in the range 10$^{22}$--10$^{24}$ cm$^{-2}$, and at least one object,
NGC 1068, with $N_{\rm H} \simgt 10^{25}$ cm$^{-2}$ (e.g. Mulchaey et al.
1992).
If these values were representative of the typical torus thickness,
most Sey 2 should be unobscured above, say, 20 keV,
and should have the same luminosity and spectrum of Sey 1
in the energy range accessible to the instrument OSSE
on the Compton Gamma Ray Observatory.
On the contrary, if the torus were typically Compton thick
($N_{\rm H} \simgt 10^{24}$ cm$^{-2}$), the flux should
be significantly reduced also at these energies.

Madau, Ghisellini \& Fabian (1993), developing the idea of Setti
\& Woltjer (1989; see also Grindlay \& Luke 1990 and Morisawa et al. 1990)
pointed out that if the mean column thickness
is of the order of $N_{\rm H}\sim 10^{24}$--$10^{24.5}$ cm$^{-2}$,
then it is possible to explain the origin of the X--ray background (XRB)
between 2 and 100 keV as emission from Sey 2 and Sey 1.
These columns are slightly larger than those typically observed by
{\it Ginga}, but the sample is likely to be biased by a selection effect,
since the brightest Sey 2 are the least absorbed.
Matt \& Fabian (1993) extended the previous analysis to include
the features introduced by the
iron $K{\alpha}$ line emission in the XRB
spectrum, and also their best models favour large column densities.
Moreover, the XRB above 10 keV would be overproduced by the Sey 2 if
the tori were transparent at these energies, provided that the
half-opening angle is $\sim$30$^{\circ}$ (e.g. Mulchaey et al. 1992).

There are therefore indications that the obscuring torus surrounding
the nucleus is thick to Compton scattering.
As a consequence, there is the need for a detailed model
of the contribution that the torus
can give to the spectrum of Sey 1 by scattering part of the
incident primary  radiation into the line of sight.
This radiation could in principle mimic the reflection bump observed in
the majority of Sey 1 (Nandra \& Pounds 1993 and references
therein), which is generally interpreted as due to reflection from
the accretion disc (Lightman \& White 1988; George \& Fabian
1991; Matt, Perola \& Piro 1991, hereafter MPP).
The strongest proof that the reflected component must come
from regions located near the primary X--ray source was the almost simultaneous
variability of the continuum and the iron line in NGC 6814,
but this is most likely due to the presence, in the field of view,
of a variable galactic source (Madjeski et al. 1993).

If the torus is Compton thick, Sey 2
should be dimmer than Sey 1 even at high X--ray energies,
because most of the high energies photons are Compton downscattered, and can
eventually be absorbed or emerge at lower energies, possibly out of the
line of sight.
This of course bears important consequences for the detectability
of Sey 2 by OSSE.
At very large energies, the decline of the Klein--Nishina cross section
allows an increasing fraction of primary photons to pass the torus unscattered.
Not only the total flux, but also its shape will therefore be a function
of the optical depth of the torus.

In this paper we calculate, by means of Montecarlo simulations, the spectrum
expected in the Seyfert unification scheme, for both Sey 1 and Sey 2,
as a function of the torus column density.
In section 2 we give details of the assumed geometry, the angular
dependence of the incident spectrum, and the Montecarlo code we use.
In Section 3 we show the results, concerning the X--ray spectrum
as a function of the inclination and of the column density, and the behaviour
of the fluorescent iron line. Finally, in section 4, we discuss our results.

First Monte Carlo results on this subject have been presented by
Awaki et al. (1991).
During the completion of this work, we became aware of similar
works in progress on the subject by Krolik, Madau \& Zycki (1993), which
independently confirms some of our results.

\vskip 1 true cm
\noindent
{\bf 2. THE MODEL}
\vskip 0.7 true cm

\noindent
In Fig. 1 we sketch the adopted geometry: the primary
X--ray source is located close to a (cold) accretion disc,
and is assumed to emit a spectrum

$$
F(\nu)\, \propto \nu^{-\alpha}\exp(-h\nu/kT) .
\eqno(1)
$$

\noindent
This spectral form has been chosen accordingly to SIGMA and
OSSE observations of Sey 1
(e.g. Jourdain et al. 1992; Maisack et al. 1993; Cameron et al. 1993;
Fabian et al. 1993) and it is a rough approximation of unsaturated
Comptonization in a thermal plasma. The e-fold energy $kT$ is only indicative
of the temperature.
We have used $kT=100$ keV, according to the typical OSSE best--fit results.
The index $\alpha$ is taken to be 0.9, consistent with the observations
of Sey 1 below 20 keV (Pounds et al. 1990; Matsuoka et al. 1990;
Nandra \& Pounds 1993).
It is worth noting that the inclusion of the cutoff  in the primary
spectrum has a negligible effect on the fluorescent iron line intensity
(see next section), as the photons effective in producing the line
emission are those with energy less than $\sim$50 keV.

We assume that the primary X--ray source is isotropic.
About half of its radiation
impinges onto the accretion disc, where it is partly reflected
by Compton scattering and fluorescence emission (Lightman \& White 1988;
George \& Fabian 1991; MPP).
The shape of the Compton reflection continuum
has been calculated following the prescriptions of White, Lightman \&
Zdziarski (1988) and Lightman \& White (1988).
The two--stream approximation
adopted by these authors allows only to compute the reflected component
integrated over the emission angles.
Unfortunately, the actual angular behaviour
of the Compton hump is complex, and it is a
function of the photon energy (George \& Fabian 1991; MPP).
Only when the
scattering can be treated as elastic an analytical estimation can be made.
For simplicity, we have used the two--stream approximation
results for the shape, multiplied by an angular function
to normalize the spectrum at different viewing angles.
In such a way, the shape of the Compton reflection
spectrum we adopted is of course independent of the emission angle.
For $E>20$ keV this is a crude approximation, but sufficient for our scope.
The angular dependence of the X--ray radiation reprocessed
by the disc is approximated by the normalized function:

$$
f(\mu)\, =\, {3\mu\over 4}
\left[ (3-2\mu^2+3\mu^4)\, \ln\left(1+{1\over\mu} \right)
+\left( 3\mu^2-1\right) \left( {1\over 2} -\mu \right)\right] ,
\eqno(2)
$$

\noindent
where $\mu$ is the cosine of the inclination angle $i$.
This formula comes from assuming a single scattering process governed by
the Rayleigh phase function in a semi-infinite medium.
It is found to be in
excellent agreement with Montecarlo simulations for energies below 15-20 keV.

{}From now on, we call `nuclear' the spectrum resulting from the
sum of the primary isotropic X--ray source (emitting the spectrum of Eq. 1)
and the reflected component from the disc.
For simplicity we have not included, in the latter component,
the iron line produced by the disc.
For a semi--isotropic illumination, the line equivalent width
(EW) is about 150 eV for a disc seen face--on (e.g. Matt et al. 1992),
and decreases with $\mu$ with a law which
can be approximated (if the matter is neutral and general relativistic
effects are ignored) by the simple formula:

$$
{\rm EW}(\mu) \, \simeq \, {150 \over \ln 2}\, \mu \ln
\left(1+ {1 \over \mu}\right)
\quad {\rm eV}
\eqno(3)
$$

\noindent
(Basko 1978; Haardt 1993).
A fraction of the line emitted by the disc can be reflected by the
funnel of the torus; we have evaluated that this effect enhances the
EW by no more than about 10 per cent at small inclination angles.

The equatorial plane of the torus has been assumed to be equal to that
of the accretion disc.
We call $R$ the distance between the primary source
(coincident with the geometrical center of the system)
and the outer walls of the
torus, and $r$ the distance between the center and the inner wall
of the torus in the equatorial plane.
We always used $r/R=0.1$.
The half opening angle of the torus, $\theta$, is defined as shown
in Fig. 1.

We indicate with $N_{\rm H}$ the column densities measured
in the equatorial plane $(i=90^{\circ})$.
At a given inclination angle $i>\theta$, the column density along the
line of sight is

$$
N_{\rm H}(i)\, =\, {1-k(i)~r/R \over {1-r/R}}~N_{\rm H} ,
\eqno(4.a)
$$
where
$$
k(i)={\cos \theta \over{ (r/R-\sin \theta)\cos i +
\cos\theta \sin i}} .
\eqno(4.b)
$$

We have assumed interstellar abundances as in Morrison \& McCammon (1983).
Further details of the adopted
atomic data as well as the radiative transfer treatment can be found in MPP.

The Montecarlo method is a on/off type.
Weights are used to model the input
spectrum, increasing the statistics in the high energy bins.
For each run we typically used 2$\times$10$^7$ photons,  while the
output is split into twenty angular bins,
equally spaced in $\mu$.

\vskip 1 true cm
\noindent
{\bf 3. RESULTS}
\vskip 0.7 true cm
\noindent
{\bf 3.1 Spectra for small viewing angles}
\vskip 0.7 true cm
\noindent
{\it 3.1.1 Continuum}
\vskip 0.7 true cm

\noindent
In Fig. 2 we show the spectra calculated for different
$N_{\rm H}$ at the viewing angle $i=0^{\circ}-18^{\circ}$,
i.e. smaller than the torus opening angle.
In this case
we see directly the Sey 1 nucleus plus the contribution from
the radiation scattered by the torus into the line of sight.
The warm scattering material
may further complicate the spectrum by absorbing and scattering the
incident radiation out of the line of sight.
Indeed, absorption by warm material seems now to be a common
feature in Sey 1 (see Nandra \& Pounds 1993 and references therein).
Hereinafter, we call WSM the warm scattering material,
and WSC the corresponding scattered continuum.
The spectral signature of the WSM for small viewing angles
has been qualitatively described by Krolik \& Kallman (1987).

For $N_{\rm H} \ll 10^{24}$ cm$^{-2}$, the radiation passing through the torus
either escapes or is absorbed, while the scattering is negligible
($\tau_{\rm T} \ll 1$) and therefore the torus does not contribute
significantly
to the spectrum.
Instead, for $N_{\rm H} \simgt 10^{24}$ cm$^{-2}$, when
the scattering optical depth
is of order unity or greater, an important fraction of the hard X--rays is
Compton scattered and reflected by the funnel of the torus,
while softer photons are photoabsorbed
(Compton scattering and photoabsorption cross section
are equal at about 10 keV for neutral matter).
Note that the Klein--Nishina decline of the scattering cross section
can make the torus optically thin for photons with the highest energies.
Increasing $N_{\rm H}$ there is an increasing
contribution, mainly at 10--50 keV, from the funnel, which
saturates when the optical depth becomes much larger than unity.
Therefore, the spectra in Fig. 2 for $N_{\rm H}>10^{25}$ cm$^{-2}$ overlap.

Note that the spectral shape of the scattered radiation is
very similar to that produced by a flat disc, as in both cases the shape
is mainly dictated by the atomic processes.
With one major difference: here photons can scatter more than once onto
the surface of the funnel, when $\tau_{\rm T}>1$ (see MPP for a discussion
of this effect in geometrically thick accretion discs).
At 30 keV, the relative contribution of the torus to the flux
is 29 and 55 per cent for $N_{\rm H}=10^{24}$ and $10^{25}$ cm$^{-2}$,
respectively.

These results have the important consequence to introduce an energy--dependent
variability.
In the soft X--rays we see only the compact illuminating source,
and a rapid variability is expected.
At larger ($\simgt$ 10 keV) energies and for large column densities
the flux from the compact source is diluted by radiation coming from
the torus, with a much larger emitting area.
Hence the latter fraction of the flux varies with a longer time--scale.
Therefore both the amplitude and the time--scale of variability
should be frequency dependent, with softer energies being
more violently variable.
As a consequence, the spectral shape should also vary,
pivoting around 20--30 keV.

Of course, similar energy--dependent variability is expected from
reflection off an accretion disc.
However, in this case the bulk of the reprocessed radiation
is expected to come from the inner 50-100 gravitational radii (MPP).
So the variability time--scale should be
of order hours--days, while the variability time--scale of the radiation
from the torus should be of order months--years (the location of the inner
surface of the torus is
very poorly determined; the constraint that the torus must be located outside
the Broad Line Region and inside the Narrow Line Region
permits to range
between a few light--months  and several light--years).
Therefore, variability
measurements are fundamental in determining the relative amount of reprocessed
radiation from the two regions.

\vskip 0.7 true cm
\noindent
{\it 3.1.2  Iron fluorescent line}
\vskip 0.7 true cm

\noindent
In Fig. 3 we show the equivalent width of the iron fluorescent line emitted
by the torus vs $N_{\rm H}$ for two values of the half-opening angle
($\theta=30^{\circ}$, left hand panel, and $\theta=45^{\circ}$, right
hand panel) and for three different angular bins
(0$^{\circ}$--18$^{\circ}$, filled squares,
60$^{\circ}$--63$^{\circ}$, open circles,
81$^{\circ}$--84$^{\circ}$, crosses).

The results for small
viewing angles indicate that the line EW is very small for
$N_{\rm H} \simlt 10^{23}$ cm$^{-2}$, but increases to
about 70-90 eV for $N_{\rm H} \simgt 5\times 10^{23}$ cm$^{-2}$.
For very thick tori
($N_{\rm H} \simgt 5\times 10^{25}$ cm$^{-2}$),
the equivalent width is greater for the 45$^{\circ}$
case because the smaller solid angle subtended by the torus
is overcompensated by a
greater projected area (see MPP for a detailed discussion of this effect).
The question which naturally arises is if this component alone
can account for the observed iron lines in Sey 1.
The mean equivalent width for the sample
studied by Nandra \& Pounds (1993) (27 sources, mainly Sey 1 but including
7 Narrow Emission Line Galaxies, hereafter NELG)
is about 110 eV (fits with the reflection
component included; the average EW is
140 eV fitting the continuum with a single power law),
but much greater and much lower values are also present.
Assuming a dispersion in the iron abundance and its mean value
somewhat greater than the solar, the observed results can be explained
by the emission from the torus, provided that the average column density
is $\simgt 10^{24}$ cm$^{-2}$.

On the other hand, the observed iron line properties are also in good
agreement with the accretion disc reflection hypothesis
(e.g. Matt et al. 1992; Nandra \& Pounds 1993).
The EW is about 150 eV for a face--on disc, and smaller for greater
viewing angles (see eq. 3), to give an angle--averaged
(between 0$^{\circ}$ and 90$^{\circ}$) value of $\sim$110 eV.
If the unification scheme is correct,
the range of visible angles appropriate for Sey 1
is limited by the aperture of the torus, and
the predicted mean value of the EW is consequently larger than the
observed 110 eV; but the presence of a few NELG in the sample and a possible
slight iron underabundance could maybe explain the discrepancy.
If the torus is Compton thick, however,
it contributes significantly to the iron line,
bringing the total average EW to about 250 eV, much greater than observed.
This argues in favour of small ($N_{\rm H} \simlt 10^{23}$ cm$^{-2}$)
column densities,
in disagreement with the estimation based on the XRB (but in agreement
with observations of an UV--selected sample, Mulchaey et al. 1992).
This apparent contradiction can in principle be solved in many ways,
for instance by assuming that
the iron abundance is smaller than the solar in both the accretion disc
and the torus, or that the contribution of one or both the components is
reduced in some way (for instance by an anisotropic primary X--ray emission
which privileges the small inclinations, or a very sharp inner
surface of the torus with consequently small projected area).
All these solutions appear rather unsatisfactory, leading to the
suggestion that if the accretion disc emits (exists ?)
the torus must be Compton thin (or to not exist at all)
while if the torus is Compton thick the disc contribution must be
small or absent.
Variability and/or high energy resolution measurements
are required to solve this problem, as the line emitted by the torus is narrow
while that emitted by the disc should be broad and with a well--defined
profile due to relativistic effects
(e.g. Fabian et al. 1989; Matt et al. 1992; Matt, Perola \& Stella 1993
for a Schwarzschild black hole, and Laor 1991; Kojima 1991 for a Kerr
black hole), and should vary with a much smaller time--scale.

While the ultimate solution of this problem seems demanded to the kind
of observations just outlined,
a deeper look inside the Nandra \& Pounds sample could give some first
indications.
The line of sight column density of the 7 NELG is between
4$\times$10$^{21}$ and 10$^{23}$ cm$^{-2}$, greater than that in our
own galaxy, and greater than the mean value of the Sey 1 in the sample
(which is another confirmation of the unification scheme).
The values
are not particularly high, but this is not surprising, as the sample has
been chosen on the basis of the X--ray brightness.
With these column densities,
the predicted EW from the torus is very small at all inclination angles
(see Fig. 3),
while the observed EWs in these sources range between 70 and 440 eV.
The error bars are too large to
allow a more detailed analysis; nevertheless, it seems quite clear
that another source of iron line photons besides the torus
does exist in these objects.

To end this discussion on the iron line emission, it must be recalled that
iron fluorescence (or recombination) photons can be
emitted by the WSM outside the torus.
However, this component, which can dominate the line emission for
X--ray absorbed sources, is not very important at small
viewing angles, when the nuclear radiation is directly seen.
For opening angles such as those considered here, the EW is in fact at
most a few tens of eV (Krolik \& Kallman 1987).

\vskip 1 true cm
\noindent
{\bf 3.2 Spectra for large viewing angles}
\vskip 0.7 true cm
\noindent
{\it 3.2.1 Continuum}
\vskip 0.7 true cm

\noindent
In Fig. 4 we show the spectra, for different
column densities, corresponding to the viewing angle
$i=60^{\circ}-63^{\circ}$, for a half-opening angle of the
torus $\theta=30^\circ$.

Below 10 keV, the spectra are the sum of two
components: 1) the radiation transmitted through the torus,
and 2) the radiation reflected by that part of the torus inner
surface which is directly visible by the observer.
For $N_{\rm H}\simlt 10^{24}$ cm$^{-2}$ the transmitted component
dominates, as indicated by the exponential shape of the spectrum.
Viceversa, the reflected component, with a corresponding
power law spectrum, dominates for
$N_{\rm H}\simgt 10^{26}$ cm$^{-2}$, when the torus is
optically thick to both photoabsorption and scattering; the amount of
reflected component depends only on the inclination angle.
For intermediate values of the column density, the reflected part
dominates at energies up to a few keV, while the transmitted part
dominates at higher energies; their ratio depends on $i$.

At high ($>$ 20 keV) energies, the spectra for
$N_{\rm H}\simlt 10^{23}$ cm$^{-2}$ are practically not modified,
while for $N_{\rm H}\simgt 10^{24}$ cm$^{-2}$ even hard
X--rays are scattered away
from the line of sight, and softened by Compton recoil.
At energies greater than about 50 keV the Klein--Nishina decline begins
to be important, lowering the optical depth. The spectra (above 20 keV)
of the radiation passing through the torus without any
scattering are shown in Fig. 5 as dashed lines,
for $N_{\rm H}=10^{24}$ and $10^{25}$ cm$^{-2}$.
These has been calculated by simply multiplying the nuclear
radiation with the factor
$\exp [-\tau_{\rm T}(i)\sigma_{{\rm KN}}(\nu)/\sigma_{\rm T}]$
where $\sigma_{\rm T}$ and $\sigma_{\rm KN}(\nu)$ are the Thomson
and Klein--Nishina cross sections, respectively.
For $i=60^\circ$ and $\theta=30^\circ$ the Thomson optical depths
are $\tau_{\rm T}(60^\circ)=0.62$ and 6.2, respectively (cf. eqs. 4).
We can see that if the torus is moderately thick ($\tau_{\rm T}\sim 1$),
the unscattered flux can be a substantial fraction of the total,
being (for $N_{\rm H}=10^{24}$ cm$^{-2}$) the $\sim$50
per cent between 20 and 50 keV, and more
at higher energies, where the scattering cross section is smaller.
In the case of $N_{\rm H}=10^{25}$ cm$^{-2}$ the main contribution to the
total flux comes from (down-)scattered photons up to 400 keV, beyond
which the unscattered radiation dominates.

In Fig. 4 the spectra are shown without any contribution from the WSM.
If present, this material would intercept the nuclear radiation emitted
at small inclination angles:
the continuous lines in Fig. 5 take into account this
contribution by adding the 0.5 per cent (see next section)
of the radiation emitted along the axis of the torus (consisting of
the nuclear component plus the contribution from the torus itself,
as shown in Fig. 2).
The spectra shown in Fig. 5 refer to the cases with
$N_{\rm H}=10^{24}$, $10^{25}$ and $10^{26}$ cm$^{-2}$.
For lower values of the column density the WSC is unimportant
at all energies.
The opening angle of the torus and the inclination are the same as Fig. 4.

The WSC can be the only observable one
at energies less than a few keV when $N_{\rm H}\simgt 10^{24}$ cm$^{-2}$,
because the photons impinging onto the torus are photoabsorbed.
There is marginal evidence for this in recent ROSAT observations.
Both Mulchaey et al. (1993)
and Turner, Urry \& Mushotzky (1993) find that the continuum shapes
of Sey 2 below the photoelectric cutoff are identical to those
of Sey 1 (cf. Turner, George \& Mushotzky, 1993).
For larger column densities the WSC becomes important
also at higher energies, diluting the radiation emerging from the torus.
For $N_{\rm H}\simgt 10^{25}$ cm$^{-2}$, the  relative importance of
the WSC dilution depends on the portion of the funnel
that is visible at a given inclination angle, and therefore on the
inclination angle itself.

The sum of different contributions results in a complex
variability pattern.

i) For $N_{\rm H} \simlt 10^{23}$ cm$^{-2}$, at all energies
photons which are scattered
by both the WSM medium and the torus should give
a negligible contribution to the total flux.
Consequently, the variability is simply that of the nuclear
source at all frequencies.

ii) For $N_{\rm H}\sim 10^{24}$ cm$^{-2}$ the dominant contribution
to the total flux above 20 keV comes from unscattered photons,
while below 10 keV the WSM can dilute the observed continuum.
Therefore in this case hard X--rays should vary more rapidly than
the soft X--rays, and with a larger amplitude.
This is just the opposite behaviour than at small
viewing angles.

iii) For $N_{\rm H}\simgt 10^{25}$ cm$^{-2}$ the flux is always dominated by
scattered photons, either by the WSM (at low
energies) or by the surface of the funnel (at high energies).
The flux should therefore be steady at all but the highest
frequencies, where the decreased scattering cross section allows
photons to pass the torus unscattered.

\vskip 0.7 true cm
\noindent
{\it 3.2.2  Iron fluorescent line}
\vskip 0.7 true cm

\noindent
In Fig. 3 the EW of the iron line produced by the torus is shown
vs the column density
for two inclination angles greater than the half--opening angle, namely
60$^{\circ}$--63$^{\circ}$ and 81$^{\circ}$--84$^{\circ}$.
As the nuclear radiation is now intercepted by the torus,
the equivalent width can be very large (more than 1 keV), provided that
the column density is $\simgt$10$^{24}$ cm$^{-2}$.

As discussed above for the continuum, the total torus emission is composed
by the line photons emitted by the visible part of the inner surface
of the torus and those escaping from its outer surface.
The EW of the two components is different:
that of the transmitted component is the greatest for
$N_{\rm H} \simgt$ a few $\times 10^{24}$ cm$^{-2}$.
Both the equivalent widths (but not the luminosities)
are almost independent of the inclination angle.
Therefore, for intermediate column densities ($\sim$10$^{24.5\div 25}$
cm$^{-2}$), the total EW
increases with $i$, as the reflected/transmitted ratio diminishes for
geometrical reasons.
For lower (when the transmitted component dominates)
and for greater (when the reflected component dominates)
column densities the EW is almost independent of the inclination.

Of course, the EW shown in Fig. 3 can be reduced by dilution with the WSC.
The WSM can also produce iron lines
by fluorescence and/or recombination processes (Krolik \& Kallman 1987)
or by resonant scattering of the continuum (Band et al. 1990).
These lines should have a total EW with respect to the
WSC of order 1 keV, and their energies should
be mainly around 6.7--6.9 keV.

\vskip 0.7 true cm
\noindent
{\it 3.2.3  The case of NGC~1068}
\vskip 0.7 true cm

\noindent
The best known Sey 2 for which the continuum
below 10 keV is likely to be dominated by the WSC is NGC~1068.
It is important to note that BBXRT (Marshall et al. 1993)
has observed in its spectrum not only the line components
expected from the WSM, but also a 6.4 keV component,
with an EW of about 1 keV, which is therefore emitted by low--ionization
iron (this is confirmed by preliminary ASCA results, Tanaka 1993).
It is interesting to investigate if this component can originate in
the torus itself.

In NGC~1068 there is no indication of intrinsic absorption in the {\it Ginga}
band, and there is no compelling evidence of a flattening of the continuum at
high energies (but see below). This constrains the column density to
be  $\simgt$10$^{25}$ cm$^{-2}$.
We seek a solution in which the 6.4 keV iron line
is not completely swamped by the WSC,
even if it dominates the continuum emission up to 10 keV.

The WSC is, in the first approximation, equal
to the emitted flux multiplied by the factor
$f=\tau_{\rm sc}\Delta \Omega/4\pi$,
where $\tau_{\rm sc}$ is the Thomson depth of the WSM and
$\Delta \Omega$ the solid angle defined by the ionization cone.

In order to constrain the amount of WSC and
the inclination angle we summarize here our arguments:

\item{1)} Although the EW has a relatively weak dependence on the inclination
angle for the column densities under consideration,
its absolute luminosity can change by orders of magnitude (smaller
luminosities at higher inclinations),
because the dominant component is that reflected by the inner
torus surface (the transmitted flux has in fact
a much weaker dependence on $i$).

\item{2)} The amount ($f$) of dilution due to the WSC which is required
to obtain the observed EW also decreases with the inclination.

\item{3)} Since $\tau_{\rm es}\propto f$, an independent estimate
of $\tau_{\rm es}$ could constrain $i$.

\item{4)} Such an estimate can be provided by identifying
the WSM with the warm absorber
seen in Sey 1.

For the following estimate, we assume a half-opening angle
of 30$^{\circ}$, deriving $\Delta \Omega/4\pi= 0.067$
(this assumes that only one `hemisphere' of the torus is visible).

According to our results, in order to have an EW equal to
1 keV in the total spectrum, $f$ must be equal to about 0.014,
4$\times 10^{-3}$ and $10^{-3}$ for
$i$= 35$^{\circ}$, 55$^{\circ}$ and 83$^{\circ}$, respectively, if
$N_{\rm H}$=10$^{25}$ cm$^{-2}$.
These numbers become 0.014, 2.5$\times 10^{-3}$ and 5$\times 10^{-4}$
if $N_{\rm H}$=10$^{26}$ cm$^{-2}$
(for larger column densities the numbers saturate).
Therefore, $\tau_{\rm sc}$ is equal to about 0.2 if the viewing
angle is just greater than the half-opening angle,
between $\sim$0.04 and $\sim$0.06 for $i$=55$^{\circ}$,
and between $\sim$0.0075 and $\sim$0.015 if the source is seen almost edge-on.
In Fig. 6 we show the spectra (without the WSC) corresponding to
$\theta=30^\circ$ and $N_{\rm H}=10^{25}$ and $10^{26}$ cm$^{-2}$
for the two extreme
chosen viewing angles (i.e. 32$^{\circ}$--$37^{\circ}$
and 81$^{\circ}$--$84^{\circ}$)
and for the 0$^{\circ}$--18$^{\circ}$ case.
A fraction $f$ of the face--on spectrum forms the WSC.

We can now constrain the inclination angle by
identifying the WSM with the warm absorber seen in
several Sey 1 (see Nandra \& Pounds 1992; Nandra et al. 1993; Fiore et al.
1993; Turner et al. 1993 and Pounds et al. 1993 for ROSAT results).
Both ROSAT and {\it Ginga} observations (see Nandra \& Pounds 1993 for the
latter) indicate that warm absorbers are common and that the
typical equivalent scattering optical depth ranges between about 0.007 and
0.07 (such values are in agreement  with indirect
evaluations of the optical depth of the WSC in this and other sources,
see e.g. Krolik \& Begelman 1988; Miller, Goodrich \& Mathews
1991; Mulchaey  et al. 1993).
This then suggests a viewing angle for NGC~1068 greater
than $\sim$50$^{\circ}$.

A different estimation of $\tau_{\rm sc}$ can be derived by an energy budget
argument. The intrinsic luminosity of NGC~1068 in the 2--10 keV band
is of course a factor
$1/f$ larger than what we see.
Assuming isotropic emission, such a luminosity is $3\times 10^{41}$
erg s$^{-1}$ (Koyama et al. 1989), and the
range in $f$ derived from the warm absorber observations
implies an intrinsic 2--10 keV luminosity of
0.6--6 $\times 10^{44}$ erg s$^{-1}$.
Integrating up to 1 MeV with our assumed spectrum gives another
factor of $\sim 2.5$.
Inclusion of a likely optical--UV component with
luminosity similar to the X--ray one brings the total
nuclear luminosity to a value of 0.3--3$\times 10^{45}$ erg s$^{-1}$,
a range of values which includes that of
the measured (if isotropic) infrared luminosity, i.e.
$L_{\rm IR}\sim 7 \times 10^{44}$ erg s$^{-1}$ (as estimated by
Miller et al. 1991 for the core component).
Assuming the latter to be entirely due to the reprocessing of the nuclear
radiation, the two spectral components must be of the same order, which
gives an $f$ of about $2.5\times 10^{-3}$
(that means $\tau_{\rm sc} \sim 0.04$) corresponding to an inclination
angle of about 55$^{\circ}$--70$^{\circ}$.

Possible problems with this result are the following:

\item{1)} If $N_{\rm H}\simeq 10^{25}$ cm$^{-2}$, the transmitted radiation
(whose luminosity is only weakly dependent on the inclination angle)
gives an important contribution above 10 keV, with a corresponding flattening
of the total spectrum. The flattening
is predicted to be smaller for lower inclination
angles, because of the increasing contribution of the radiation
reflected by the funnel, whose flattening is less dramatic than that of
the transmitted component (see Fig. 5).
Note that for $N_{\rm H} \simgt 10^{26}$ cm$^{-2}$ the
transmitted component vanishes.
We estimate that after dilution with the WSC,
the spectrum in this case would resemble
a typical Sey 1 spectrum, but with a ratio between the
normalizations of the
Compton reflection component and of the power law much greater
(e.g. about seven times if $f$=0.014, corresponding to $i$=35$^{\circ}$)
than that for a 2$\pi$--illuminated slab.
The presence of such a component is allowed (even if not required)
in the spectral fitting of NGC~1068 done by Smith, Done \& Pounds (1993).
It is worth noting that
with such a strong reflection component, the power law continuum
becomes closer to the canonical value ($\alpha_{2-10}=0.9$) of Sey 1.

\item{2)} A much more serious problem in having large inclinations
may be the observed 16 per cent degree of polarization of the broad lines.
A very simple calculation, based on the formulae of Brown \&
McLean (1977), gives an inclination angle of about 35$^{\circ}$
for $\theta$=30$^{\circ}$. Even
if the WSM is located just above the torus,
the viewing angle is not greater than 38$^{\circ}$
(a value of 60$^{\circ}$, consistent with our estimation,
would give a polarization of 48 and 37 percent in the two cases).
Of course, it is well possible that the actual optical depth
of the WSM in NGC~1068 is unusually high, and that therefore we have
overestimated the inclination angle; the energy budget argument can be
readjusted by assuming an anisotropic emission of the nuclear and/or
the IR radiation, or a much different bolometric correction.
However, the other Sey 2 for which broad lines has been
detected in scattered light have a much smaller degree of polarization
(Miller \& Goodrich 1990),
which is very surprising if NGC~1068 were really seen at an angle
just greater than the minimum value allowed for a Sey 2.
One solution could be an optically
thick WSM, so that the polarization could be diminished by multiple
scatterings, but we have no evidence for a Thomson thick absorber in
Sey 1. Another possible solution, i.e. much greater half-opening angles,
contrasts with the measurements of the ionization cones.
Of course, more detailed models
of the polarization are required, but one cannot help wondering if we
are missing an ingredient or grossly misunderstanding the geometry
[however, there are recent indications that in some Sey 2 the broad lines
are much more polarized than the continuum (Miller 1993), reaching values
greater than 20 \%.
This would resolve the polarization problem at the expense
of introducing an extended source of featureless continuum with the
same spectrum of the nuclear one, as the polarization is in general
wavelength--independent.
Of course, this could imply that NGC 1068 is really see at low inclination
and that we are overestimating this angle.
A more precise determination of the EW of the cold iron line component,
which will surely be obtained by ASCA, will allow a more reliable
determination of the viewing angle].

\vskip 0.7 true cm
\noindent
{\bf 3.3  Polarization}
\vskip 0.7 true cm

\noindent
In principle, another way to distinguish between disc and torus
reprocessed emission and to constrain the torus thickness is by
polarization measurements. In fact, a certain amount of polarization of
the torus scattered emission is expected, because the geometry is not
spherical.
The degree of polarization of the total radiation should depend strongly
on the inclination angle, due to the
angular dependence of the polarization of the
radiation scattered by the torus and the amount
of visible nuclear radiation (which is assumed to be unpolarized).

In Fig. 7 we show the degree of polarization of the torus emission
as a function of $\mu$
for $N_{\rm H}$=10$^{24}$ cm$^{-2}$ (left hand panel) and
$N_{\rm H}$=10$^{25}$ cm$^{-2}$ (right hand panel) for
three different
energy ranges: 2--5.6 keV (squares), 16--45 (circles)
and 126--355 keV (crosses).
All results refer to the case of a torus with half-opening angle of
$30^\circ$. By convention, the polarization is negative when
aligned with the projection of the torus axis onto the plane of the sky,
and positive when parallel to it (for symmetry reasons
these are the only two allowed directions of the total polarization).

Let us discuss firstly the case $N_{\rm H}$=10$^{24}$ cm$^{-2}$. In this
case, the torus emission is in general
a mixture of the reflected and transmitted components. It must be noted that
in general the reflected radiation is more polarized than the transmitted one;
furthermore, the former is positive while the latter is negative.

In the lowest energy bin the torus is opaque, and therefore
the reflected component dominates. As a result, the
polarization is positive and rather high (of order 5-10 per
cent), provided that the
nucleus is hidden. Note that the polarization increases with $\mu$, due
to the increasing ratio between the reflected and transmitted
components. The degree of polarization decreases suddenly when
the nucleus becomes visible, because now the (scattered) photons from
the torus are diluted by the direct emission (besides the fact that the
polarization of scattered radiation itself diminishes with $\mu$).

At higher energies the transmitted component dominates, and therefore
the polarization is negative, and with a lower degree. The highest energy
bin has a lowest degree of polarization than the intermediate one because
at those energies the cross section is lower, and more nuclear (unpolarized)
photons contribute, and the scattering is less efficient in
polarizing radiation (McMaster 1961). The polarization is practically
zero for the Sey 1.

If $N_{\rm H}$=10$^{25}$ cm$^{-2}$ the reflected component is the most
important at all
energies, and the polarization is always positive. At the lowest energy bin
it is very high for the Sey 2, because there is no contribution from the
nuclear radiation, completely absorbed by the torus. At greater
energies the transmitted component is significant, albeit not dominant, and
dilutes the reflected, highly polarized radiation. At
these energies the degree of polarization increases with $\mu$ following
the increasing ratio of the reflected to the transmitted radiation.
Also for this column density the polarization is very small for the Sey 1
at all energies.

Unfortunately, by comparing Fig. 7 with the previous figures, it can be easily
seen that in general the polarization is higher when the flux is smaller
(this is indeed a rather general problem with polarization).
This reduces very much the actual possibility to use polarization
as a check of the model.

In principle, polarization measurements in Sey 1
can be used to test the simplest version of the unification model
by exploiting the fact that the accretion disc reflected component
is also expected to be polarized (always negative) with a strong angular
dependence (Matt 1993).
In fact, if the torus exists in all Sey 1, such sources should be seen
at low inclinations, and the polarization of the reflected component is
expected to be small. On the contrary, if relatively high degrees of
polarization (a few percent at energies $\simgt$10 keV) are observed,
this argues in favour of high inclinations and therefore of absence
(or very large opening angle) of the torus.

Of course, if the WSM is composed by electrons (as seems to be the case
for NGC~1068), a positive polarization with a degree
very similar to that of the broad lines is expected. As the relative
importance of the WSC and the torus emission depends on the energy,
this introduce a further frequency dependence in the polarization properties.

Finally it should be considered that the primary X-rays could be
intrinsically polarized, for instance if the emission is due to
thermal Comptonization in a non spherical corona
(Sunyaev \& Titarchuk 1985). In this case, the polarization is expected to
be a strong function of the photon energy (Haardt \& Matt 1993).

\vskip 1 true cm
\noindent
{\bf 4. SUMMARY AND DISCUSSION}
\vskip 0.7 true cm

 \noindent
We have investigated how the X--ray spectra of Seyfert galaxies
are modified by absorption and scattering of the primary source in the
molecular torus surrounding the nucleus.
Our results can be summarized as follows:

\item{1)} At small inclination angles, we have shown that
Compton thick tori can significantly contribute to the emission at 10--50 keV,
resembling the Compton reflection hump produced by cold
accretion discs in the vicinity of the primary X--ray source.

\item{2)} The torus, for small inclination angles and
for $N_{\rm H}\simgt 10^{24}$ cm$^{-2}$,
can produce an iron $K{\alpha}$ line with EW of $\sim 90$ keV, not much
smaller than the average value observed in Seyfert galaxies.
This indicates that a thick torus alone can produce the observed
EWs, but in this case the iron line from the disc must be dimmed in some way.
Otherwise, the torus must be in general Compton thin.

\item{3)} The presence of the torus and the WSM introduces
a complex pattern for the predicted variability, which can be used
to estimate the relative strength of the different components.
For $N_{\rm H}\simgt 10^{24}$ cm$^{-2}$, the soft X--rays
of Seyfert 1 should vary more rapidly than the hard X--rays.
The spectrum should pivot around 20--30 keV.
The opposite behavior is predicted in Seyfert 2
for intermediate values of the column density
($N_{\rm H}\sim 10^{24}$ cm$^{-2}$).
For larger column densities the X--rays should not vary on short time--scales
in all bands.

\item{4)} The torus produces a substantial amount of iron 6.4 keV line
even at large inclinations. The EW with respect to the
continuum emerging from the torus is of order of a few keV,
as long as $N_{\rm H}\simgt 10^{25}$ cm$^{-2}$.

\item{5)} In our model, the identification of
the warm scattering material with the warm
absorber observed in Seyfert 1 suggests that
NGC 1068 is viewed at an angle greater than about 50$^{\circ}$.

Some of the results outlined before deserve a more extended discussion.

For small viewing angles, when the observer sees directly the nucleus,
the equivalent width of the iron line emitted by the torus is very small
for $N_{\rm H} \simlt$10$^{23}$ cm$^{-2}$ and of order to 70-90 eV for
$N_{\rm H} \simgt$5$\times$10$^{23}$ cm$^{-2}$. The accretion disc
can also emit a
fluorescent line, with EW up to about 150 eV for a face--on disc (e.g.
Matt et al. 1992).
A comparison with the observed values (Nandra \& Pounds 1993) suggests
that the two contributions can hardly be present simultaneously; the presence
of the accretion disc requires a Compton thin torus, while if the torus is
Compton thick, the accretion disc (if present) can not contribute
significantly. The results for the seven NELG included in the Nandra \&
Pounds sample strongly suggest that a line emitting region besides
the torus does exist. Anyway,
the two contributions can, in principle, be distinguished by
their width (the line emitted by the torus must be much narrower than that
emitted by the disc) and by their variability time--scales (that
of the disc emission being much smaller than the one of the torus emission).
{\it Ginga} measurements of the line width are inconclusive (Nandra \&
Pounds 1993), but much firmer constraints should be put by forthcoming
ASCA measurements.
With regard to the variability, it is worth noting that
the recent discovery that the claimed periodicity in the X--ray emission
of NGC~6814 is actually due to a galactic source in the field of view
(Madjeski et al. 1993) implies that the most compelling reason to locate
the line emitting matter very close to the black hole (Kunieda et al. 1990)
is now probably gone away. Nevertheless,  it is worth noting that at least
another, even if less spectacular,
evidence for short time--scales of the reprocessed radiation
does exist (NGC~4051, Fiore et al. 1992).

An interesting result of our computations is that
the torus can easily account for the cold iron line component observed
in NGC 1068 (Marshall et al. 1993), provided that the column density
in this source is of order 10$^{25}$--10$^{26}$ cm$^{-2}$, as
indeed suggested by {\it Ginga} and BBXRT observations
(Koyama et al. 1989; Marshall et al. 1993; Smith et al. 1993).
In fact, the iron line
in the transmitted spectra has equivalent widths of order a few
keV, which can diminishes to the observed $\sim$ 1 keV after dilution with
the scattered nuclear radiation.
Assuming for the optical depth of the
WSM, $\tau_{\rm sc}$, values consistent with the observations of
warm absorbers in Seyfert 1 (e.g. Nandra \& Pounds 1993), we are able to
say that this source is probably seen at a viewing angle greater than
$\sim$50$^{\circ}$, while an independent estimation of $\tau_{\rm sc}$
based on an energy budget argument suggests $i \sim$55$^{\circ}$--70$^{\circ}$.
However, such
range of inclinations is inconsistent with the value derived from
the polarization degree of the broad lines, which is 35$^{\circ}$, a
value derived by assuming the simplest geometry and an optically thin case.
Of course, our evaluations are based on rather indirect evidence, and
$i$ can be grossly overestimated.
However, the inclination derived from the polarization
argument is close to the minimum value for which the nucleus is
obscured, so that the other Seyfert 2
should show in general higher polarization,
while the contrary is actually observed (Miller \& Goodrich 1990).

As discussed above, one way to test our results is by measuring in detail
the iron line profile. Such a task should be within the capabilities of
ASCA, at least for the brightest sources. Another way to test the model
is by broad--band temporal measurements, as our calculations predict
energy--dependent variability. Missions like XTE and SAX, which will
cover a wide energy range, should enlighten us on the relation between
soft and hard X--rays. Unfortunately, polarization measurements, which
would provide an invaluable tool to study the geometry of the reprocessing
medium,
require sensitivities beyond those of the planned X--ray polarimeters
(but possibly not beyond the capabilities of the present generation
of Thomson scattering polarimeters in a mission completely
devoted to X--ray polarimetry).

Finally, our results can have important consequences on recent models
of the XRB, in which a significant contribution
is made by heavily obscured Seyfert 2 galaxies,
as firstly suggested by Setti \& Woltjer (1989). In these models
the column density of the torus required to fit the XRB
is $\simgt 10^{24}$ cm$^{-2}$ (Madau et al. 1993), or a distribution
of values with a significant contribution from sources with
$10^{24} < N_{\rm H} < 10^{25}$ cm$^{-2}$ (Matt \& Fabian 1993).
These values are somewhat in excess of the constraints
posed by our model and by the observations of the EW of the iron line
in Seyfert galaxies, which indicate $N_{\rm H}\simlt 10^{23}$ cm$^{-2}$.
However, this disagreement is probably not serious, and can be possibly
reconciled by exploring a wider range of the space of parameters
(Madau \& Ghisellini 1993)
and by considering possible selection effects in the present samples
of Seyfert galaxies, which favours the closest Seyfert 1 and the least
obscured (and therefore brightest) Seyfert 2.

\vskip 1 true cm
\noindent
{\bf ACKNOWLEDGMENTS}
\vskip 0.5 true cm

\noindent
FH acknowledges financial support from Italian MURST, GM from an E.S.A.
fellowship.

\vfill\par\eject

\baselineskip=14pt

\vskip 1 true cm
\noindent
{\bf REFERENCES}
\vskip 0.5 true cm

\parindent=0 pt
\everypar={\hangindent=2.6pc}

Antonucci R.R., 1993, ARA\&A, in press

Antonucci R.R., Miller J.S., 1985, ApJ, 297, 621

Awaki H., Koyama K., Inoue H., Halpern J.P., 1991, PASJ, 43, 195

Band D.L., Klein R.I., Castor J.I., Nash J.K., 1990, ApJ, 262, 90

Basko M.M., 1978, ApJ, 223, 268

Brown J.C., McLean I.S., 1977, A\&A, 57, 141

Cameron R.A., et al., 1993, Proceedings of the Compton symposium, Eds.
N. Gehrels, M. Friedlander, N. Gehrels, D.J. Macomb, St. Louis, p. 478

Fabian A.C., Rees M.J., Stella L., White N.E., 1989, MNRAS, 238, 729

Fabian A.C., Nandra K., Celotti A., Rees M.J., Grove J.E.,
Johnson W.N., 1993, ApJ, 416, L57

Fiore F., Perola G.C., Matsuoka, M., Yamauchi M., Piro L.
1992, A\&A, 262, 37

Fiore F., Elvis M., Mathur S., Wilkes B., McDowell J., 1993, ApJ, 415, 129

George I.M., Fabian A.C., 1991, MNRAS, 249, 352

Grindlay J.E., Luke M., 1990, in IAU Colloq. 115, eds. P.
Gorestein, M. Zombeck, Dordrecht:Kluwer, p. 276

Haardt F., 1993, ApJ, 413, 680


Haardt F., Matt G., 1993, MNRAS, 261, 346

Haniff C.A., Ward M.J,, Wilson A.S., 1991, ApJ, 368, 167

Jourdain E., et al. 1992, 256, L38

Kojima Y., 1991, MNRAS, 20, 629

Koyama K., Inoue H., Tanaka Y., Awaki H., Takano S., Ohashi
T., Matsuoka M. 1989, PASJ, 41, 731

Kunieda H., Turner T.J., Awaki H., Koyama K., Mushotzky R.F.,
Tsusaka Y., 1990, Nat, 345, 786

Krolik J.H., Begelman M.C., 1988, ApJ, 329, 702

Krolik J.H., Kallman T.R., 1987, ApJ, 320, L5

Krolik J.H., Madau P., Zycki P., 1993, ApJ, submitted

Laor A., 1991, ApJ, 376, 90

Lawrence A., Elvis M., 1982, ApJ, 256, 410

Lightman A.P., White T.R, 1988, ApJ, 335, 57

Madau P., Ghisellini G., Fabian A.C, 1993, ApJ, 410, L10

Madau P., Ghisellini G., 1993, in preparation

Madjeski G., et al., 1993, Nat, in press

Maisack M.,  et al., 1993, ApJ, 407, L61

Marshall F.E., et al., 1993, ApJ, 405, 168

Matsuoka  M., Piro  L., Yamauchi M., Murakami T. 1990, ApJ, 361, 440

Matt G., 1993, MNRAS, 260, 663

Matt G., Fabian A.C., 1993, MNRAS, in press

Matt G., Perola G.C., Piro L., 1991, A\&A, 247, 25 (MPP)

Matt G., Perola G.C., Stella L., 1993, A\&A, 267, 643

Matt G., Perola G.C., Piro L., Stella L., 1992, A\&A, 257, 63
(Erratum in A\&A, 263, 453)

McMaster W.H., 1961, Rev. Mod. Phys., 33, 8

Miller J.S., Goodrich R.W., 1990, ApJ,  355, 456

Miller J.S., Goodrich R.W., Mathews W.G., 1991, ApJ, 378, 47

Miller J.S., 1993, in "The Physics of Active Galactic Nuclei", Canberra,
July 1993, in press

Morisawa K., Matsuoka M., Takahara F.,  Piro L., 1990, A\&A, 236, 299

Morrison R., McCammon D., 1983, ApJ, 270, 119

Mulchaey J.S., Mushotzky R.F., Weaver K.A., 1992, ApJ, 390, L69

Mulchaey, J.S., Colbert E., Wilson A.S., Mushotzky R.F., Weaver K.A.,
1993, ApJ, 414, 144

Nandra K., Pounds K.A., 1992, Nat, 359, 215

Nandra K., Pounds K.A., 1993, submitted to MNRAS

Nandra K., et al., 1993, MNRAS, 260, 504

Pogge R.W., 1989, ApJ, 345, 730

Pounds K.A., Nandra K., Stewart G.C., George I.M.,
Fabian A.C., 1990, Nat, 344, 132

Pounds K.A., Nandra K., Fink H.H., Makino F., 1993, MNRAS, in press

Setti G., Woltjer L., 1989, A\&A, 224, L21

Smith D.A., Done C., Pounds K.A., 1993, MNRAS, 263, 54

Sunyaev R. A., Titarchuk L. G., 1985, A\&A, 143, 374

Tadhunter, C., Tsvetanov, Z., 1989, Nat, 341, 422

Tanaka Y., 1993, in Active Galactic Nuclei across the Electromagnetic
Spectrum, IAU Symposium 159, in press

Tran H.D., Miller J.S., Kay L.E., 1992, ApJ, 397, 452

Turner T.J., Nandra K., George I.M., Fabian A.C., Pounds K.A., 1993, ApJ,
in press

Turner T.J., Urry C.M., Mushotzky, R.F., 1993, ApJ, in press

Turner T.J., George, I.M., Mushotzky, R.F., 1993, ApJ, 412, 72

White T.R., Lightman A.P., Zdziarski A., 1988, ApJ, 331, 939

Wilson A.S., 1992, in Physics of Active Galactic Nuclei,
eds. W.J. Duschl, S.J. Wagner, (Springer--Verlag), p. 307

Wilson A.S., Elvis M., Lawrence A., Bland--Hawthorn J. 1992, ApJ,
391, L75

\vfill\par\eject

\vskip 1 true cm
\noindent
{\bf FIGURE CAPTIONS}
\vskip 0.5 true cm

{\bf Figure 1}: The geometry assumed for our computations.
The source of primary X--ray radiation is located at the center
of an obscuring torus with column density $N_{\rm H}$ along the
equatorial plane.
This source is assumed isotropic.
A cold accretion disc (not shown in the figure)
is assumed to intercept half of the primary
X--rays and to contribute to the primary spectrum by Compton reflection.
The angular dependence of this component is given by eq. 4.
The torus semiaperture angle is $\theta$, and the viewing angle
is $i$.
In our calculation, we always used $r/R=0.1$.

{\bf Figure 2}: Spectra computed with different values of $N_{\rm H}$
(increasing from bottom to top)
for an inclination angle $0^\circ$--$18^\circ$.
The assumed torus semi--aperture angle is $\theta=30^\circ$.

{\bf Figure 3}: The equivalent width of the fluorescence iron line
produced by the torus as a function of $N_{\rm H}$ for two
values of the half-opening angle
($\theta=30^{\circ}$, left hand panel,
and $\theta=45^{\circ}$, right hand panel),
and for three different angular bins
(0$^{\circ}$--18$^{\circ}$, filled squares,
60$^{\circ}$--63$^{\circ}$, open circles, and 81$^{\circ}$--84$^{\circ}$,
crosses).

{\bf Figure 4}: Spectra computed with different values of $N_{\rm H}$
(increasing from top to bottom)
for an inclination angle $60^\circ$--$63^\circ$.
The assumed torus half--opening angle is $\theta=30^\circ$.

{\bf Figure 5}: Histograms: spectra for $N_{\rm H}=10^{24}$,
$N_{\rm H}=10^{25}$ and $N_{\rm H}=10^{26}$ cm$^{-2}$ (from top to bottom),
assuming $\theta=30^\circ$ and an inclination angle $60^\circ$--$63^\circ$.
Continuous line: to the calculated spectra, we have added 0.5 \%
of the nuclear radiation (emitted at $i=10^\circ$)
to mimic the possible contribution of the WSM.
The dashed lines corresponds to the nuclear radiation emitted
at $60^\circ$ multiplied by the factor
$\exp[-\tau_{\rm T}(60^\circ)\sigma_{\rm KN}(\nu)/\sigma_{\rm T}]$,
for the $N_{\rm H}=10^{24}$ and $N_{\rm H}=10^{25}$ cm$^{-2}$ cases
(top to bottom). These lines indicate the radiation passing through the torus
without any scattering.

{\bf Figure 6}: The upper (overlapping) histograms
are the spectra observed at $i=0^\circ$--$18^\circ$
for $N_{\rm H}=10^{25}$ and $N_{\rm H}=10^{26}$ cm$^{-2}$.
The other histograms are the spectra viewed at
$i=32^\circ$--$37^\circ$ and $81^\circ$--$84^\circ$,
for the same column densities
(increasing from top to bottom),
appropriate for the Sey 2 galaxy NGC 1068.
A fraction $f$ of the upper curves can dilute the shown radiation
emerging from the torus at large inclinations.
Note that the amount of dilution, $f$, necessary to hide the
rising spectrum and to produce an iron line EW of 1 keV depends
on the inclination.

{\bf Figure 7}: The degree of polarization as a function of $\mu$, the cosine
of the inclination angle, for $\theta=30^\circ$ and two values of the
column density, namely $N_{\rm H}=10^{25}$ (left hand panel) and
for $N_{\rm H}=10^{26}$ cm$^{-2}$. Three different energy bins are
shown: 2-5.6 keV (filled squares), 16-45 keV (open circles) and 126-355 keV
(crosses).

\bye